# Extended Ensemble Molecular Dynamics for Thermodynamics of Phases

Gang Seob Jung[1,†], Yoshihide Yoshimoto[2], Kwang Jin Oh[3], Shinji Tsuneyuki[4,5,†]

[1]Computational Sciences and Engineering Division, Oak Ridge National Laboratory, Oak Ridge, TN 37831, USA
[2]Graduate School of Information Science and Technology, The University of Tokyo, Japan
[3]Center for Supercomputing Technology Development, Supercomputer Division, Korean Institute of Science and Technology Information, South Korea
[4]Department of Physics, The University of Tokyo, Japan
[5]Institute for Solid State Physics, The University of Tokyo, Japan

[†]Email: jungg@ornl.gov, stsune@phys.s.u-tokyo.ac.jp

**Abstract**: The first-order phase transitions and related thermodynamics properties are primary concerns of materials sciences and engineering. In traditional atomistic simulations, the phase transitions and the estimation of their thermodynamic properties are challenging tasks because the trajectories get trapped in local minima close to the initial states. In this study, we investigate various extended ensemble molecular dynamics (MD) methods based on the multicanonical ensemble method using the Wang-Landau (WL) approach. We performed multibaric-multithermal (MBMT) method to fluid phase, gas-liquid transition, and liquid-solid transition of the Lennard-Jones (LJ) system. The derived thermodynamic properties of the fluid phase and the gas-liquid transition from the MBMT agree well with the previously reported equation of states (EOSs). However, the MBMT cannot correctly predict the liquid-solid transition. The multiorder-multithermal (MOMT) ensemble shows significantly enhanced sampling between liquid and solid states with an accurate estimation of transition temperatures. We further investigated the dynamics of each system based on their free energy shapes, providing fundamental insights for their sampling behaviors. This study guides the prediction of broader crystalline materials, e.g., alloys, for their phases and thermodynamic properties from atomistic modeling.

Thermodynamics properties and phases of materials are the most fundamental concerns in materials design and development. While electronic structure-based approaches such as density functional theory (DFT) have been successfully applied to predict material properties from molecular or atomistic structures, identifying thermodynamic behaviors across different phases generally requires sampling approaches. Monte Carlo (MC) and molecular dynamics (MD) methods combined with DFT, or empirical interatomic potentials, have played an essential role in understanding the thermodynamics or phase transition of materials[1-11] using generalized/extended ensemble methods, including multicanonical ensemble and the replica-exchange method. These methods are designed to overcome the local energy-minimum problem where sampling is more likely trapped in metastable states.

In the multicanonical ensemble method, a non-Boltzmann weight factor is used to sample wide ranges of energy or enthalpy for a given system. The multicanonical ensemble can be extended, referred to as extended ensemble methods, using the two-dimensional weight factors.[7,8,12,13] Since the non-Boltzmann factor is not given a *priori*, several iterations should be performed to determine the weight factor[3]. This conventional approach can be inefficient, depending on the systems. The Wang-Landau (WL) algorithm has successfully improved the sampling efficiency in MC method.[4] Also, previous studies show successful sampling with the WL-based MD simulations.[7,8]

This study systematically compares the various isothermal-isobaric ensemble MD schemes and investigates how they behave when applying the WL algorithm. The results show that the WL process is inapplicable to some commonly utilized isothermal-isobaric ensemble schemes due to the numerical instability of the time integrator. Nosé-Hoover chain-based NPT ensemble proposed by Martyna, Tobias, and Klein (MTK)[14] can work successfully with the WL process. Then, we perform the multibaric-multithermal (MBMT) and multiorder-multithermal (MOMT) ensemble MDs for various phases of the Lennard-Jones (LJ) systems and compare the thermodynamic quantity with the previously reported values. Furthermore, we compare the obtained non-Boltzmann factors from different systems and analyze how they govern sampling behaviors and their efficiency.

## 2. Methods
### 2.1. *Multicanonical ensemble molecular dynamics*
Here, we briefly review the formulations of the extended ensemble methods from the multicanonical ensemble methods[3,7,15]. In the canonical ensemble, the probability distribution at the temperature $T_0$ is given by

$$P_{NVT}(E) = \frac{1}{Z_{NVT}} n(E) e^{-E/k_B T_0} \quad , \tag{1}$$

where $n(E)$ is the states; $Z_{NVT} = \sum n(E) e^{-E/k_B T_0}$; $E$ is energy; $e^{-E/k_B T_0}$ is a Boltzmann factor for NVT ensemble. The distribution is bell-shaped because $n(E)$ increases, and the Boltzmann factor decreases as $E$ increases. Therefore, it is difficult to sample both high- and low-energy regions from the canonical ensemble, as shown in Figure 1a.

Figure 1b shows how the multicanonical ensemble method can overcome the inefficient sampling problem of the canonical ensemble. In the multicanonical ensemble method, a weight factor, $W_{mc}(E)$, is utilized for the artificially flat probability distribution as

$$P_{mc}(E) = \frac{1}{Z_{mc}} n(E) W_{mc}(E) = const., \tag{2}$$

where $Z_{mc} = \sum n(E) W_{mc}(E)$. We can estimate $W(E)_{mc}$ from a canonical distribution ($P_c$) at a given temperature ($T_0$) as $W_{mc} \sim \frac{1}{n(E)} \sim \frac{e^{-\frac{E}{k_B T_0}}}{P_c} = e^{-\frac{E + kT_0 \ln P_c(E, T_0)}{k_B T_0}} = e^{-\frac{E_{mc}(E)}{k_B T_0}}$.

We then can consider $E_{mc}(E)$ as a deformed potential as $E_{mc}(E) = E + kT_0 \ln P_c(E, T_0)$[3]. The multicanonical potential, $E_{mc}(E)$, is not given a *priori*, and we should obtain and expand it through iterations with a constant temperature $T_0$:

$$E_{mc}^{i+1}(E) = E_{mc}^i(E) + k_B T_0 \ln P_{mc}(E). \tag{3}$$

Figure 1c represents a schematic of the potential scaling method showing how the deformed potential can sample a wide range of energy surfaces with constant kinetic energy in MD simulations. The equations of motion are governed by modified forces by the chain rule,

$$\boldsymbol{f}_i^{mc} = -\frac{\partial E_{mc}}{\partial r_i} = -\frac{dE_{mc}}{dE}\frac{\partial E}{\partial r_i} = -\frac{dE_{mc}}{dE}\boldsymbol{f}_0, \tag{4}$$

where $r_i$ and $f_0$ are coordinates and forces from the original potentials.

One of the advantages of utilizing the multicanonical ensemble is the reweighting technique[16] to obtain canonical distribution $P_{NVT}$ from $P_{mc}$ and $W_{mc}$ as

$$P_{NVT}(E) = \frac{P_{mc}(E) W_{mc}(E) e^{-E/k_B T_0}}{\int P_{mc}(E) W_{mc}(E) e^{-E/k_B T_0} dE}. \tag{5}$$

Once we obtain enough random sampling of the energy space, we can obtain an ensemble-averaged thermodynamic quantity at any temperature.

### 2.2. Wang-Landau algorithm in molecular dynamics

Wang et al. have devised an algorithm in MC to estimate the density of states with high accuracy and efficiency.[4] In the algorithm, the transition probability from the energy $E_1$ to $E_2$ is given by

$$p(E_1 \to E_2) = \min\left[\frac{g(E_1)}{g(E_2)}, 1\right], \tag{6}$$

where the $g(E)$ is the density of states. It searches the energy space with a probability proportional to $1/g(E)$ and can efficiently search the states. The algorithm modifies the density of states from 1 by multiplying a Wang-Landau (WL) factor, $w_i>1$, when the states are visited. Therefore, it realizes the random work on the energy spaces and quickly searches the possible energy space. After the accumulated histogram is flat enough, the next iteration starts with a smaller factor, $w_{i+1} = \sqrt{w_i}$, to more accurately refine the densities of states. We deform the potential energy surface in the MD by adding a Gaussian-like biased potential. This process is equivalent to the metadynamics[17] in Figure 1d.

Now, multicanonical potential in MD simulations is expressed with partial energy $\delta E$ as
$$E_{mc} = E_0 + \delta E(E), \tag{7}$$
where $\delta E(E)$ is a biased potential accumulated by the Gaussian-like biased potential; $E_0$ is the energy from the original potential. $\delta E(E)$ is estimated in an on-fly manner during the simulations. When the histogram becomes flat, $\delta E$ is modified by a smaller WL factor $w_{i+1} = 0.5\, w_i$ in the next iteration. The same process is repeated iteratively until the WL factor ($w_{final}$) becomes small enough. Then, the production run is performed based on determined $\delta E$ and $w_{final}$[7]. Ideally, a zero factor is desired as the final WL factor in the production run, which is identical to the conventional approach. However, this can be achieved in only limited cases with MD. We will investigate the conditions in more detail later.

*2.3. Isobaric-multithermal ensemble molecular dynamics*
From the distribution of the isobaric-isothermal (IBIT) ensemble,
$$P_{NPT}(H) = \frac{1}{N_{NPT}} n(H) e^{-H/k_B T_0}, \tag{8}$$
where enthalpy, $H = E + P_0 V$; $V$ is volume; $P_0$ is pressure. From the same concept in equation (7), the multicanonical enthalpy can be expressed by
$$H_{mc}(H) = H_0 + \delta H(H), \tag{9}$$
where $H_0$ is enthalpy from the original potential and a given pressure $P_0$; $\delta H(H)$ is the partial enthalpy and is updated during simulations. Like multicanonical forces in eq (3), the partial enthalpy $\delta H$ is a function of $H$. The forces, virial pressure, and external pressure are given from the chain rule[7,13],
$$\boldsymbol{f}_0 \rightarrow \boldsymbol{f}_{mc} = -\frac{\partial H_{mc}}{\partial r_i} = \left(1 + \frac{\partial \delta H(H)}{\partial H}\bigg|_{H_c}\right) \boldsymbol{f}_0, \tag{10}$$
$$\frac{1}{3V} \sum_{i=1}^{N} \boldsymbol{r}_i \cdot \boldsymbol{f}_0 \rightarrow \frac{1}{3V} \left(1 + \frac{\partial \delta H(H)}{\partial H}\bigg|_{H_c}\right) \sum_{i=1}^{N} \boldsymbol{r}_i \cdot \boldsymbol{f}_0, \tag{11}$$
$$P_0 \rightarrow \frac{\partial H_{mc}}{\partial V}\bigg|_{H_c} = P_0 \left(1 + \frac{\partial \delta H(H)}{\partial H}\bigg|_{H_c}\right), \tag{12}$$
where $H_c$ is the current enthalpy, and $\boldsymbol{f}_0$ is the original forces. The reweighting technique from the factor, $W_{mc} = e^{-\frac{H_{mc}(H)}{k_B T_0}}$, can derive the distribution of isobaric-isothermal ensemble:
$$P_{NPT} = \frac{P_{mc} W_{mc} e^{-H/k_B T_0}}{\int P_{mc} W_{mc} e^{-H/k_B T_0} dH}. \tag{13}$$

Here, we consider the meaning of $\delta H(H)$ in more detail. The entropy of a system, $S(H)$, is defined as $k_B \ln n(H)$, where $n(H)$ is the density of states that can be approximated as
$$n(H) \sim W_{mc}^{-1} = e^{H_{mc}(H)} \tag{14}$$
Gibbs free energy at the temperature $T_0$ is given by,
$$G = E + P_0 V - T_0 S \sim E + P_0 V - T_0 k_B \ln e^{H_{mc}(H)} = -\delta H(H) \tag{15}$$
Therefore, we can consider the relative Gibbs free energy ($\Delta G$, not absolute free energy) as $-\delta H$.

*2.4 Multibaric multithermal ensemble molecular dynamics*
In the multibaric multithermal (MBMT) formulations, the partial enthalpy is now a function of energy ($E$) and volume ($V$) as
$$H_{mbmt} = H_0 + \delta H(E, V). \tag{16}$$
The forces, virial pressure, and external pressure are given by
$$\boldsymbol{f}_0 \rightarrow \boldsymbol{f}_{mbmt} = -\frac{\partial H_{mbmt}}{\partial r_i} = \left(1 + \frac{\partial \delta H(E,V)}{\partial E}\bigg|_{E_c, V_c}\right) \boldsymbol{f}_0, \tag{17}$$
$$\frac{1}{3V} \sum_{i=1}^{N} \boldsymbol{r}_i \cdot \boldsymbol{f}_0 \rightarrow \frac{1}{3V} \left(1 + \frac{\partial \delta H(E,V)}{\partial E}\bigg|_{E_c, V_c}\right) \sum_{i=1}^{N} \boldsymbol{r}_i \cdot \boldsymbol{f}_0, \tag{18}$$
$$P_0 \rightarrow \frac{\partial H_{mbmt}(E,V)}{\partial V}\bigg|_{E_c, V_c} = P_0 + \frac{\partial \delta H(E,V)}{\partial V}\bigg|_{E_c, V_c}, \tag{19}$$
where $E_c$ and $V_c$ are the current states of energy and volume; $\boldsymbol{f}_0$ is the original force from the defined interatomic potentials.

*2.5 Multiorder multithermal ensemble molecular dynamics*

In the isobaric-multiorder multithermal (MOMT) ensemble method, the partial enthalpy is defined as functions of $H$ and $O$ as

$$H_{momt} = H_0 + \delta H(H,O). \tag{20}$$

where $O$ is an order parameter function to evaluate the order of the system as a function of coordinates, $r$. The derived formulations are

$$\boldsymbol{f}_0 \rightarrow \boldsymbol{f}_{momt} = -\frac{\partial H_{momt}}{\partial \boldsymbol{r}_i} = \left(1 + \frac{\partial \delta H(H,O)}{\partial H}\bigg|_{H_c,O_c}\right)\boldsymbol{f}_0 - \frac{\partial \delta H(H,O)}{\partial O}\frac{\partial O}{\partial \boldsymbol{r}}\bigg|_{H_c,O_c}, \tag{21}$$

$$\frac{1}{3V}\sum_{i=1}^{N} \boldsymbol{r}_i \cdot \boldsymbol{f}_0 \rightarrow \frac{1}{3V}\sum_{i=1}^{N} \boldsymbol{r}_i \cdot \left[\left(1 + \frac{\partial \delta H(H,O)}{\partial H}\bigg|_{H_c,O_c}\right)\boldsymbol{f}_0 - \frac{\partial \delta H(H,O)}{\partial O}\frac{\partial O}{\partial \boldsymbol{r}}\bigg|_{H_c,O_c}\right], \tag{22}$$

$$P_0 \rightarrow \frac{\partial H_{momt}(H,O)}{\partial V}\bigg|_{H_c,O_c} = P_0\left(1 + \frac{\partial \delta H(H,O)}{\partial H}\bigg|_{H_c,O_c}\right) + \frac{\partial \delta H(H,O)}{\partial O}\frac{\partial O}{\partial V}\bigg|_{H_c,O_c}, \tag{23}$$

where $H_c$ and $O_c$ are the current values of enthalpy and order parameters.

One successful order parameter to describe liquid-solid phases can be derived from the previous study[7]:

$$O = \frac{1}{N_{fcc}N_A}\sum_{g\in fcc}\left|\sum_i \exp(i g \cdot r_i)\right|^2, \tag{24}$$

where $g$, $N_{fcc}$, and $N_A$ are reciprocal vectors of FCC (depending on the volume), the number of reciprocal vectors, and the number of atoms, respectively. Since reciprocal vectors rely on the volume of the system (e.g., $\boldsymbol{b}_1 = \frac{2\pi}{V}\boldsymbol{a}_2 \times \boldsymbol{a}_3$) and three components in the virial pressure term, the second terms of eq (22) and eq (23) are canceled out in pressure balance. Therefore, the simplified formulations are derived as

$$\frac{1}{3V}\sum_{i=1}^{N} \boldsymbol{r}_i \cdot \boldsymbol{f}_0 \rightarrow \frac{1}{3V}\sum_{i=1}^{N} \boldsymbol{r}_i \cdot \boldsymbol{f}_0\left(1 + \frac{\partial \delta H(H,O)}{\partial H}\bigg|_{H_c,O_c}\right) \tag{25}$$

$$P_0 \rightarrow \frac{\partial H_{momt}(H,O)}{\partial V}\bigg|_{H_c,O_c} = P_0\left(1 + \frac{\partial \delta H(H,O)}{\partial H}\bigg|_{H_c,O_c}\right). \tag{26}$$

We tested various Hamiltonian-based isobaric isothermal ensembles (Supplementary Information) through MM_PAR molecular dynamics packages[18] by implementing the derived formulations.

### 2.6 Wang-Landau process in isothermal-isobaric molecular dynamics

Previous studies have applied the WL approach to MD simulations[7,8,19]. However, more attention must be paid to the effects of the WL process on isothermal-isobaric ensemble formulations. Compared to MC simulation, the WL process on the isobaric-isothermal molecular dynamics has a critical problem that comes from the non-conservation of the total energy. When the WL factor is zero, the dynamics are identical to the conventional multicanonical ensemble MD. A numerical instability becomes critical with the non-zero WL factor. We applied the WL algorithm to various isobaric-isothermal ensemble methods and discussed their stability (Supplementary Information). Hamiltonian-based isobaric-isothermal methods have a critical issue with the time integration of volume. Nosé-Hoover chain-based NPT ensemble proposed by Martyna, Tobias, and Klein (MTK)[14] works stably even with the massive additional energy from the WL process at every WL step ($\Delta t_{WL}$) (Figure S1).

To estimate accurate thermodynamic quantity with the WL process, the WL factor ($w_i$) should be small enough without missing states, and the partial enthalpy, $\delta H$, should not change much during the production run. However, the transition rate between the two phases (e.g., solid-liquid transition with MBMT) becomes severely lower as the WL factor decrease. To overcome this, we do not decrease the WL factor too small. Instead, we perform the production run several times with a large enough WL factor to estimate the partial enthalpy and distribution ($\delta H$, $P_{mc}$). To evaluate partial enthalpy ($\delta H$), we discretize two-dimensional spaces of energy and volume in the MBMT, and two-dimensional spaces of enthalpy and order parameter in the MOMT. The partial enthalpy, $\delta H$, is interpolated by the third-order linear polynomial[13]. Therefore, we can obtain the first derivatives utilized in the equations of motion.

As $\delta H$ gets bigger and wider, the sampling region becomes too wide, reducing the sampling efficiency. Various approaches can be considered to resolve this. For example, applying a high energy barrier can restrict the sampling region. In this study, we empirically fix the maximum value of $\delta H$ during the process to not oversample outside the region we are interested in. We can define the sampling region without additional bias by controlling the reference temperature and maximum of $\delta H$. If the maximum value of $\delta H$ is larger than the specified limit with an amount of energy ($\Delta \varepsilon$), $\delta H$ is subtracted by the same amount of $\Delta \varepsilon$ in all areas. Then, all negative values on the grids are adjusted as zero. It can lose information around the boundary of $\delta H$ but allows us to estimate thermodynamic quantities several times without losing sampling efficiency.

Another important setting is the size of the WL factor. The forces from the partial enthalpy are determined by the first derivative, *e.g.*, $\partial \delta H / \partial E$. We define the initial WL factor as $w_0 = 0.025 \Delta E$ or $0.025 \Delta H$, where $\Delta E / \Delta H$ is the space of the grid in the energy/enthalpy axis. This guarantees the maximum force modification from a WL factor, not more than 0.025, at each WL step. The initial shape of $\delta H$ determines the grid number of the 2$^{nd}$ axis as the shape looks like continuous Gaussian. We modify the first neighbor grids with a $w_0/2$ and the second neighbor grids with a $w_0/4$ in the first and second WL iterations to accelerate the sampling. The third iteration uses $w_0$ without modification in neighbor grids. The fourth and fifth iterations use $w_4 = w_0/2$ and $w_5 = w_0/4$, respectively. Then, we perform five production runs without reducing the WL factor further.

It is critical to determine how to decide the flatness of the sampling histogram for the next WL iteration. We define the region with $\delta H$ larger than the value of 0.3 $\delta H_{max}$ as effective areas on the mesh points. Then, we average the histogram. When a mesh points over a specific criterion, *e.g.*, 80% of the average of the histogram, the point is counted as a randomly sampled region. Then, we calculate the flatness as 'randomly sampled region'/'total effective areas.' We increase the flatness criterion, *e.g.*, 40% to 80%, during the five WL iterations. We reset the histogram as one and start the next WL iteration once the calculated flatness is larger than the flatness criterion.

We have applied the approach to Lennard-Jones particle systems (particle $N = 500$) for fluid phase, liquid-gas transition, and solid-liquid transition and compared the obtained quantity with previously obtained equations of states (EOSs) from different approaches, including MC simulations. We use the reduced unit for the quantities in the simulations: $T^* = k_B T / \varepsilon$ ; $P^* = P\sigma^3 / \varepsilon$ ; $E^* = E / N\varepsilon$ ; $V^* = V / N\sigma^3$ , where $\varepsilon$ and $\sigma$ are the potential depth and diameter of the Lennard-Jones potential. The detailed setting for each system is described below.

*Fluid phase*
There are 50 grids in $E^*$ (-5.4, - 3.0) and 40 grids in $V^*$ (1.11, 1.88) with $\delta H^*_{max} = 0.72$. Time step is $\Delta t = 2$fs. $R_{cut} = 3.6\sigma$ for LJ particles. Reference temperature and pressure are $T_0^* = 2.0$ and $P_0^* = 3.0$. The WL step is $\Delta t_{WL} = 40$ step.

*Liquid-gas phase*
There are 100 grids in $E^*$ (-6.57, - 0.08) and 40 grids in $V^*$ (0.5, 17.8) with $\delta H^*_{max}$ = 0.72. Time step is $\Delta t$ = 2fs. $R_{cut}$ = 3.6$\sigma$ for LJ particles. Reference temperature and pressure are $T_0^*$ =1.2 and $P_0^*$ = 0.1. The WL step is $\Delta t_{WL}$ = 200 step.

*Solid-liquid phase (MBMT)*
There are 75 grids in $E$ (-9.26, - 4.21) and 45 grids in $V$ (0.71, 1.42) with $\delta H^*_{max}$ = 0.72. Time step is $\Delta t$ = 2fs. $R_{cut}$ = 3.6$\sigma$ for LJ particles. Reference temperatures and pressures are ($T_0^*$ =0.71, $P_0^*$ = 2.38) and ($T_0^*$ =0.95, $P_0^*$ = 3.00). The WL step is $\Delta t_{WL}$ = 40 step.

*Solid-liquid phase (MOMT)*
In MOMT ensemble MD, the parameters and setting are 80 meshes in $E$, 200 meshes in $O$ with $\Delta t$ =2fs, $R_{cut}$ = 3.6$\sigma$. Different from the MBMT, we need to set different pressures to obtain the phase boundary. The details are listed in Table 1. The WL step is $\Delta t_{WL}$ = 40 step. We utilized the order parameters of FCC in eq (24), where the reciprocal vectors depend on the volume of the system.

### 3. Results and Discussion
Nosé-Hoover chain-based NPT ensemble (MTK) method[14] shows stabilized behavior during the WL process for isobaric-isothermal ensemble MD (Supporting Information). It is possible to perform a long simulation because the motion of the equations does not directly depend on the potential energy of the thermostat. Here, we demonstrate that the MBMT and MOMT ensemble methods with the suggested WL process work successfully for various phases.

*3.1 Multibaric-multithermal ensemble for fluid phase*
Figure 2a shows the comparison between the IBIT and MBMT ensemble MDs in terms of sampling in the energy and volume spaces. The MBMT with the WL process (black lines) quickly reaches the broader range of sampling than two IBIT at $T^*$=2.4 and $T^*$=1.6 with $P^*$=3.0 (See Movie 1). Figure S2 and S3 show the evolution of partial enthalpy $\delta H$ ($E^*$, $V^*$) and sampling histograms during the WL iteration from 2$^{nd}$ to 9$^{th}$. The partial enthalpy, $\delta H$, shows a very smooth profile, making the first derivative of the partial enthalpy continuous. In the previous study, MBMT MD was successfully performed for fluid phases without the WL process because of the smoothness of $\delta H$. The conventional iterative way to estimate $\delta E$ or $\delta H$ from eq (3) is nicely working.

We perform the production run from the 6$^{th}$ to the 10$^{th}$ for the thermodynamic quantities. The reweighting technique is applied. Figure 2b shows an example of canonical distribution functions at different temperatures, $T^*$=1.6, $T^*$=2.0, and $T^*$=2.4, derived from the final production run. We estimate the <E>* and <V*> from the canonical distributions and determine mean values with min/max errors from each production run. The obtained thermodynamic quantities show good agreement with equations of states (EOS) derived in the previous studies, as in Figure 2c showing the results with modified Benedict-Webb-Rubin (MBWR) EOS[20,21].

*3.2 Multibaric-multithermal ensemble for liquid-gas phase transition*
Now we move on to the next system, the liquid-gas phase transition. Figure 3a shows the time evolution of energy and volume from the IBIT (red and blue) and MBMT (black) ensemble MDs. Different from the example of the fluid phase, we select only one condition ($T^*$=1.15, $T^*$=0.061) for both IBIT and MBMT. The IBIT only samples liquid or gas phase, depending on the initial phase, due to the high energy barrier. Figure S4 and S5 show the evolution of partial enthalpy $\delta H(E^*, V^*)$ and sampling histograms during the WL iterations from 2$^{nd}$ to 9$^{th}$. We can clearly observe two local maxima in the partial enthalpy $\delta H(E^*, V^*)$. Considering the meaning of the partial enthalpy in eq (15), those are two local minima of liquid and gas phase on Gibbs free energy surface ($\Delta G$) at the reference temperature $T_0^*$=1.2.

We note that we utilize a longer WL step, $\Delta t_{WL} = 200$, than the fluid phase. From liquid to gas or gas to liquid, a relaxation time is required due to the artificial mass of volume. If we modify the potential faster than the step we define, the system becomes easily unstable in the early stage, and the volume expands too fast from liquid beyond the gas phase. We can adjust the step shorter once $\delta H(E^*, V^*)$ has an overall shape, but we apply the same WL step for the entire simulation in this study.

As expected, reweighting techniques can provide thermodynamic properties $<E>^*$, $<V^*>$, and $<H^*>$ from the reweighted canonical distributions. The heat capacities are derived from $C_p = \frac{\partial <H>}{\partial T}$ for various pressures in Figure 3c. We can clearly observe how the heat capacity changes as the pressure and temperature change. As the pressure becomes close to the critical point at $T^*$=1.31 ~ 1.33, $P^*$=0.124 ~ 0.147[12], the heat capacity decreases, and it becomes challenging to distinguish gas and liquid phases as expected. We calculated the transition temperatures from the specific heat under given pressures and compared them with the previous liquid-gas coexisting conditions in Tables 2 and 3, showing good agreement. As the reference pressure increases, we note that the transition temperatures from the heat capacity are slightly lower than coexisting temperatures. This is because the coexisting conditions are derived from the equal probability distributions of phases, while the peak of heat capacity is not necessary to be the equal probability distributions.

*3.4 Multibaric-multithermal ensemble for solid-liquid phase transition*
Next, we explore the solid-liquid phase. The time evolution of energy and volume from the IBIT (red and blue) and MBMT (black) ensemble MDs are shown in Figure 4a. The IBIT only samples solid or liquid phases at the same conditions ($T^*$=1.15, $T^*$=0.061) but with different initial phases. The MBMT with the WL process (black lines) can sample both liquid and solid phases, but the sampling behaviors are more likely distinguishable compared to fluid or liquid-gas phases. Figures S6 and S7 show the evolution of partial enthalpy $\delta H(E^*, V^*)$ and sampling histograms during the WL iterations from 2$^{nd}$ to 9$^{th}$. The partial enthalpy clearly shows two local maximums corresponding to two free energy local minimums of solid and liquid phases. Unlike the shape of the liquid-gas phase, we realize that there is a sharp boundary between the two phases. Considering eq (17), the boundary indicates that there is discontinuity of *$f_{mbmt}$* along the energy axis. This boundary moves as the WL sampling performs. We observe that the transitions from one to the other are likely to happen as this boundary entirely covers the other phase's local maxima of $\delta H$.

The discontinuity of the forces is not critical in MC simulations because MC usually does not need to calculate the forces for the sampling. As discussed in the previous study[3], the scaled forces of the multicanonical ensemble represent the thermal fluctuations induced by the multicanonical ensemble. When the scaling factor is $1 + \left.\frac{\partial \delta H(E,V)}{\partial E}\right|_{E_c,V_c} > 1$, the interactions of particles are exaggerated compared to the reference temperature while the scaling factor,

$1 + \left.\frac{\partial \delta H(E,V)}{\partial E}\right|_{E_c,V_c} < 1$, reduces the effects of interactions. Therefore, it allows the sampling of lower and higher energy regions compared to the canonical ensemble method. Interestingly, this mechanism generally does not work well for solid-liquid phase transition in MBMT MD due to the shape of $\delta H$.

We have tried to run MBMT MD at different reference temperatures and pressures, whether we can overcome this. The discontinuity of scaling forces is not simply resolved by adjusting the reference conditions while relative heights of local maxima of $\delta H$ change. Figure S8 and S9 show the evolution of partial enthalpy $\delta H$ ($E^*$, $V^*$) and sampling histograms during the WL iterations at ($T_0^*$ =0.95, $P_0^*$ = 3.00).

### 3.5 Multiorder-multithermal ensemble for solid-liquid phase transition
We can estimate the partial enthalpy $\delta H$ through the MBMT. However, there are some difficulties in accurate and efficient sampling for solid-liquid phases through the MBMT. One is the force discontinuity, as described in the previous section. The other is the complexity of two solid phases with LJ particles (FCC and HCP). We apply the MOMT ensemble MD, and our results show that the MOMT can successfully overcome the difficulties.

Figures 5a and b show the time evolution of enthalpy and crystal phases from the MBMT and MOMT MDs. Using a previously developed code, we utilize Steinhardt's bond order parameters (Q4, Q6, W4, and W6) to label the crystal phases.[22] We calculate the norm distance from the reference values of the order parameters (Table 4) to classify phases. The MBMT primarily samples HCP structures more than FCC, while the MOMT samples the FCC phase more efficiently (See Movies 2 and 3). Also, transition rates between solid and liquid are much higher with the MOMT ensemble MD. Figure 5c shows the time evolution of the order parameters labeled by phases. Based on the order parameter's definition, we expect higher values to be close to FCC crystals, and zero represents liquid or amorphous phases.

We found that HCP phases are under 200 in the $O$ axis. Figure 5d shows the heat map of - $\delta H$ ($H^*$,$O$), which represents the relative free energy of phases ($\Delta G$) at the reference temperature and pressure ($T_0^*$ =0.71, $P_0^*$ = 0.02). Figures S10 and S11 show examples of the time evolution of partial enthalpy $\delta H$ ($H^*$,$O$) and sampling histograms during the WL iterations. We note that the HCP phase is not always sampled in each iteration. However, the reweighted properties from each iteration show good agreement because we account for overall $\delta H$ ($H^*$,$O$) effects by setting the minimum $P_{mc}$ as one during the reweighting. Figures 5e and f show examples of reweighted enthalpy and heat capacity at $P_0^*$ = 1.0. We note that the values are well-matched among the WL iterations compared to the MBMT MD.

### 3.6 Partial Enthalpy
It is worth looking at the shapes of obtained partial enthalpy $\delta H$ because the values are directly related to the density of states and, therefore, an essential physical quantity, Gibb's free energy. Also, the first derivative with energy or enthalpy governs sampling dynamics in the phase spaces. Figure 6 shows the first derivative of $\delta H$ of fluid, liquid-gas, solid-liquid-MBMT, and solid-liquid-MOMT. For fluid phases, $\delta H$ is a concave function with one local maximum (Figure S2). Figure 6a indicates that the first derivative, $\left.\frac{\partial \delta H(E,V)}{\partial E}\right|_{E_c,V_c}$, is a smooth function so that the dynamics of the multicanonical ensemble is not disturbed by abrupt

changes of multicanonical force, $f_{mc} = \left(1 + \frac{\partial \delta H(E,V)}{\partial E}\bigg|_{E_c,V_c}\right) f_0$. This smooth force on the sampling space is also observed in the liquid-gas system in Figure 6b. Although the $\delta H$ of the liquid-gas phases has two local maxima and a barrier in between, the force is smooth, allowing random sampling based on the MBMT MD. However, the force is generally not smooth in solid-liquid phases because the volume and energy/enthalpy are insufficient to distinguish between solid and liquid phases. The MBMT has a significant discontinuity of the forces between solid and liquid phases in Figure 6c. During the WL sampling, the phase transition is less likely to occur with this discontinuity. Instead, it happens once the discontinuous boundary expands over the other maximum by accumulated WL factors. Therefore, the transition rate, as in Figure 5a, gets lower as the WL factor decreases.

This discontinuity problem can be resolved with MOMT MD as the order parameter provides a smooth path from the liquid to the FCC phase. Therefore, the transitions between FCC and liquid do not change much as the WL factor decreases. There is a discontinuity between HCP and liquid phases in Figure 6d. This makes long sampling once MOMT starts to explore the HCP phase. However, a narrow path between solid phases exists through the order parameter, allowing still efficient sampling compared to the MBMT.

Finally, we compare the obtained transition temperatures from MBMT and MOMT with previously reported EOS[23-25] in Figure 7 using the empirical equation
$$P_{coex} = \beta^{-5/4} \exp(-0.4759\beta^{1/2})[16.89 + A\beta + B\beta^2], \tag{27}$$
where $\beta = 1/k_B T$, and $A$ and $B$ are parameters differently fitted to the data from different methods with MC and MD methods[23-25]. The MBMT ensemble MD at high and low reference temperatures ($T_0^* = 0.71$, $P_0^* = 2.38$ and $T_0^* = 0.95$, $P_0^* = 3.00$) significantly underestimate transition temperatures with large error bars. However, the predictions from the MOMT method are well-matched with those reference lines.

## 4. Conclusions

In this study, we systematically study isobaric-isothermal ensemble molecular dynamics with the WL process for various phases and transitions. There is a significant problem with the motions of equations derived from the Hamiltonian-based Nosé thermostat/Anderson barostat, Gaussian thermostat/Anderson barostat, and Nosé -Poincare thermostat/Anderson barostat. The WL process increases the kinetic energy; therefore, the thermostat potential increases to keep the system's temperature constant. The variable of the thermostat potential increases during the time integration, and the time integration becomes numerically unstable because the variable is directly coupled with the volume equation.

Nosé Hoover chain-based NPT ensemble MD formalism can avoid this problem, although the total Hamiltonian also increases. We modify the forces and external and internal pressures from the Nosé Hoover chain isothermal and isobaric ensemble equations for an extended ensemble.

We systematically tested our formalisms with the LJ system by applying WL process-based multicanonical ensemble MD to fluid, liquid-gas, and solid-liquid phases. The reweighted thermodynamics properties from MBMT are well-matched with previously reported data for fluid phase and liquid-gas transition. However, there was a significant limitation in lowering transition rates as the WL factor decreased when we applied the MBMT to solid-liquid phases. The MOMT method can successfully overcome the limitation, and derived thermodynamic properties, especially transition temperatures, agree well with the previous studies under various pressures.

We compared the partial enthalpy obtained from the WL process and found that the discontinuity of the first derivative along the energy or enthalpy determines the efficiency of the sampling. The continuous derivative of partial enthalpy provides a smooth transition from one state to another. However, if there is a discontinuity, it is impossible to sample between phases efficiently. The MOMT method can alleviate the problem by providing smooth paths between targeted solid and liquid phases.

The discontinuity makes the random sampling require too long simulations or a big WL factor to obtain a flat distribution of sampling in two phases. We designed the reweighting technique to minimize the inefficiency from the discontinuity by initializing the histogram as one at the beginning of the WL process. Also, the errors of reweighted values are estimated in several WL processes.

The MBMT ensemble MD works well for both the gas phase and the liquid-gas phase transitions. However, it is less likely to sample the perfect FCC crystal for the solid-liquid transition with lower transition temperatures. Our results indicate that the MOMT MD outperforms the MBMT ensemble MD in solid-liquid phases by sampling the perfect FCC crystals. Also, the MOMT shows much higher rates of transition between liquid-solid phases.

Our works can be expanded to metallic systems utilizing the order parameters based on crystal structures for thermodynamic properties at the atomic scales, which can provide essential information for alloy design. Also, it is expected to have more efficient sampling for small molecules' polymorphism by defining appropriate order parameters in the future. More importantly, wide ranges of sampling in temperature and pressure would play a critical role in the recent development of machine-learning-based potentials[26].

**Supporting Information**
Supporting discussions 1, Figures S1 – S11, Movies 1-3

**ORCID**
Gang Seob Jung: 0000-0002-8047-6505

**ACKNOWLEDGEMENTS**
G.S.J. acknowledges support by the Exascale Computing Project (17-SC-20-SC) and the Laboratory Directed Research and Development (LDRD) Program of Oak Ridge National Laboratory, managed by UT-Battelle, LLC, for the US Department of Energy under contract DEAC05-00OR22725.

**Competing interests:** The authors declare that they have no competing interests. Data and materials availability: All data needed to evaluate the conclusions in the paper are present in the paper and/or the Supporting Information. Additional data related to this paper may be requested from the authors.

**Table 1.** Conditions for MOMT ensemble MD

| $T^*$ | $P^*$ | $H^*$ | $O$ |
|---|---|---|---|
| 0.71 | 0.02 | -9.26 ~ -3.37 | -2 ~ 500 |
| 0.71 | 0.2 | -9.26 ~ -2.53 | -2 ~ 500 |
| 0.71 | 0.5 | -8.42 ~ -2.53 | -2 ~ 500 |
| 0.78 | 1.0 | -8.42 ~ -2.53 | -2 ~ 500 |
| 0.78 | 2.38 | -7.16 ~ -1.26 | -2 ~ 500 |
| 1.0 | 4.0 | -5.05 ~ 0.84 | -2 ~ 500 |
| 1.2 | 6.0 | -3.79 ~ 2.11 | -2 ~ 500 |
| 1.2 | 8.0 | -1.26 ~ 4.63 | -2 ~ 500 |

**Table 2.** The liquid-gas equilibrium from the NPT plus test particle method by Lofti et al.[27] The temperature values of MBMT are obtained from transition temperatures from the specific heat between liquid and gas phases under the given pressure by the reweighting techniques.

| $T^*$ | $T^*$ (MBMT) | $P^*$ |
|---|---|---|
| 1.00 | 1.004 (0.00) | 0.02505(22) |
| 1.05 | 1.040 (0.001) | 0.03384(43) |
| 1.10 | 1.082 (0.001) | 0.04511(83) |
| 1.15 | 1.132 (0.001) | 0.05974(41) |
| 1.20 | 1.181 (0.002) | 0.07718(66) |
| 1.25 | 1.228 (0.002) | 0.0973(11) |
| 1.30 | 1.276 (0.002) | 0.1204(23) |

Table 3. Liquid-gas coexistence from work by Kofke[28]. The temperature values of MBMT are obtained from transition temperature from the specific heat between gas and liquid phases under the given pressure by the reweighting techniques.

| $T^*$ | $T^*$ (MBMT) | $P^*$ |
|---|---|---|
| 1.053 | 1.039 (0.001) | 0.0338 |
| 1.111 | 1.089 (0.001) | 0.0471 |
| 1.176 | 1.150 (0.001) | 0.0659 |
| 1.205 | 1.183 (0.001) | 0.0774 |
| 1.25 | 1.228 (0.002) | 0.097 |
| 1.299 | 1.268 (0.001) | 0.1167 |

Table 4. Steinhardt's bond order parameters for each phase.

|  | $Q4$ | $Q6$ | $W4$ | $W6$ |
|---|---|---|---|---|
| FCC | 0.19094 | 0.57452 | -0.159317 | -0.013161 |
| HCP | 0.09722 | 0.48476 | 0.134097 | -0.012442 |
| Icosahedral | 0 | 0.66332 | 0 | -0.169754 |
| Liquid | 0 | 0 | 0 | 0 |

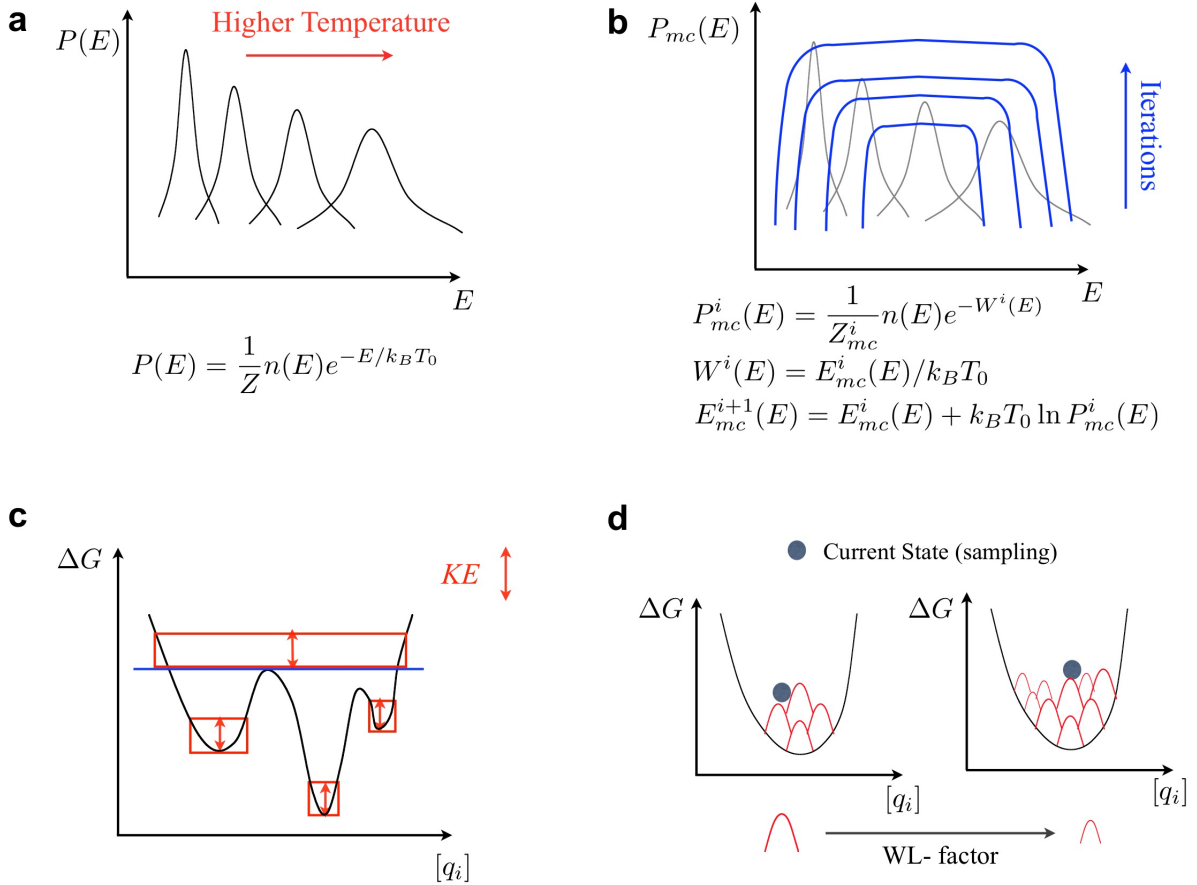

**Figure 1**. **a**. Schematics of probability distributions of canonical ensemble and **b**. multicanonical ensemble. The probability distribution of canonical ensemble shows a bell-shaped distribution while that of multicanonical ensemble shows flat distributions. Multicanonical ensemble allows us to sample a wide range of energy space, but the non-Boltzmann weight factor, $W^{i+1}(E)$, is calculated from the previous distribution, $\ln P_{mc}^{i}(E)$. **c**. Schematic of multicanonical ensemble MD with the potential scaling method[3] with a constant temperature (constant kinetic energy, KE). The black line indicates the original free energy surface. The blue line represents the deformed surface. **d**. Wang-Landau algorithm with the potential scaling method. Once sampling visits a state, the WL factor is imposed as a bias potential to lower the visiting rate for the next sampling. The WL process can be considered as metadynamics[17] by selecting energy as the reaction coordinate[9]. Multicanonical ensemble requires a flat distribution of sampling that is not mandatory in the metadynamics concept[17].

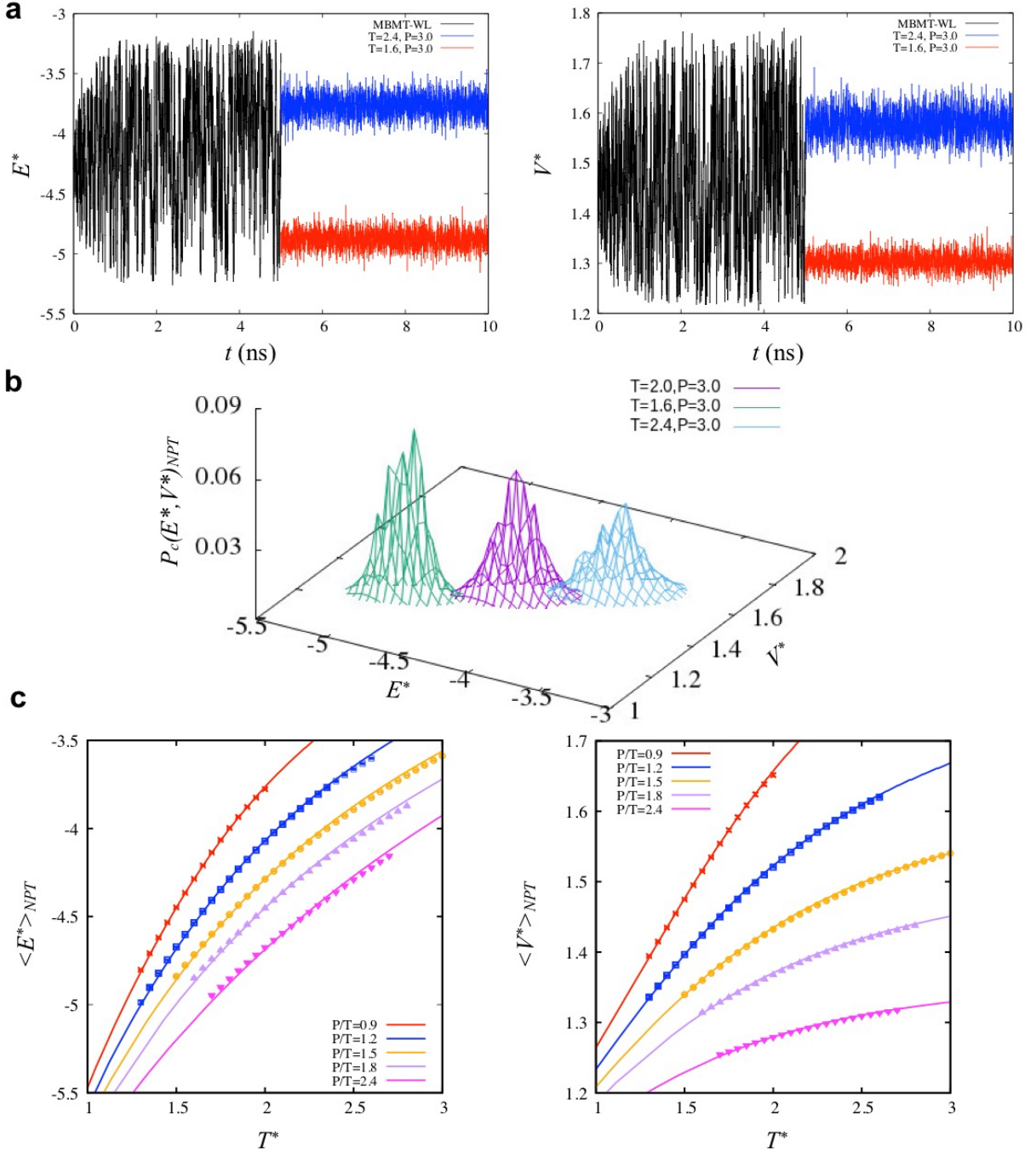

**Figure 2**. **a**. Time evolution of energy per atom ($E^*$) and volume per atom ($V^*$) of fluid system with MBMT ensemble (black: $T^*=2.0$, $P^*=3.0$) and two canonical ensembles (red: $T^*=1.6$, $P^*=3.0$ and blue: $T^*=2.4$, $P^*=3.0$). MBMT ensemble quickly broadens the sample regions to cover canonical ensembles at both temperatures. **b**. Probability distribution reweighted from the final production run of MBMT MD. **c**. $<E^*>_{NPT}$ and $<V^*>_{NPT}$ from five production runs of MBMT MD with mean values and errors. Solid curves obtained from MBWR equations[20,21].

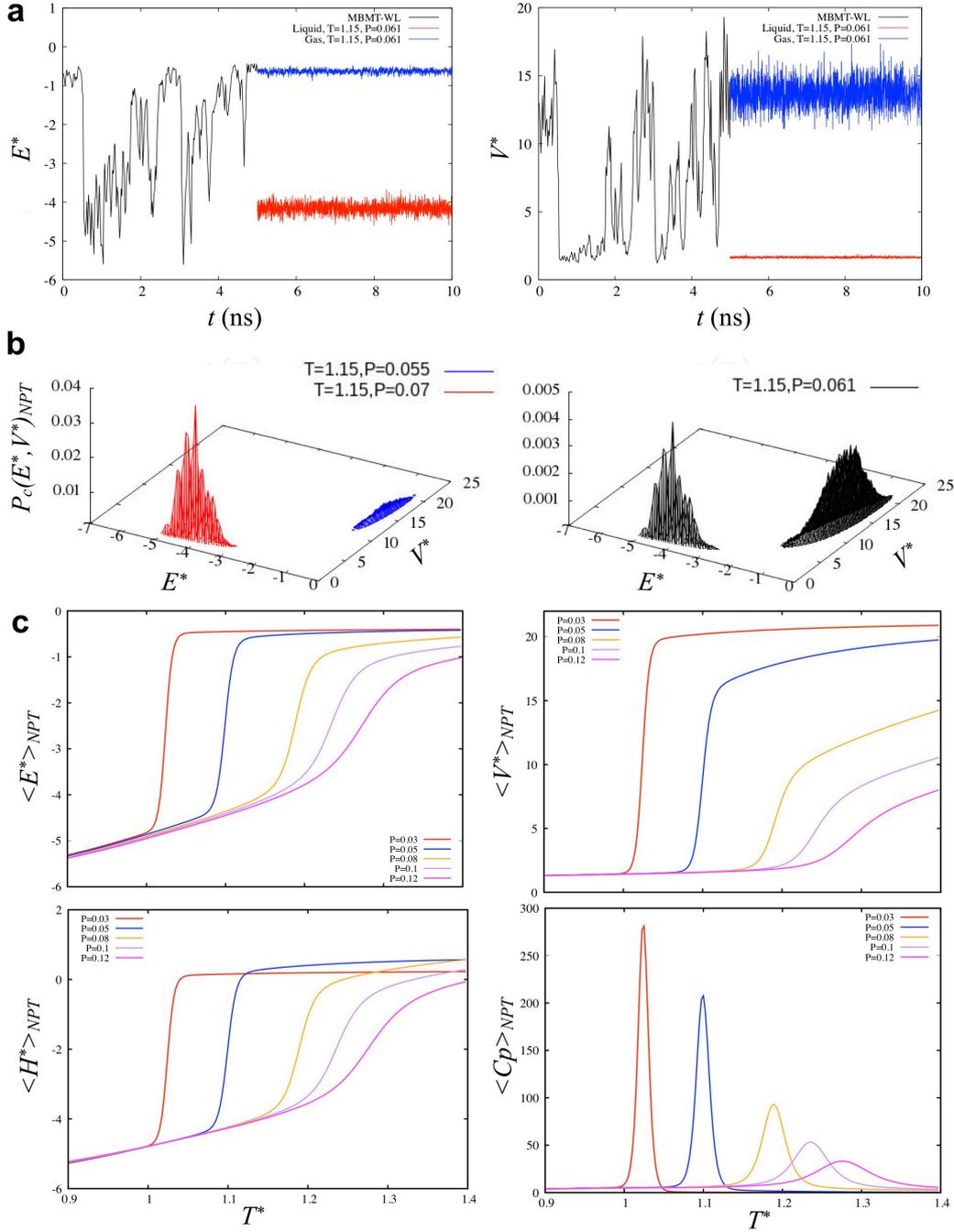

**Figure 3**. **a**. Time evolution of energy per atom ($E^*$) and volume per atom ($V^*$) of liquid-gas system with MBMT ensemble (black: $T^*$=1.15, $P^*$=0.061) and two canonical ensembles with different initial conditions at the same temperature and pressure. One starts from the liquid phase (red), and the other starts from the gas phase (blue). Due to the high energy barrier between the two phases, the IBIT ensemble cannot sample both phases, while MBMT with the WL process can sample both phases. **b**. Reweighted probability distributions from the final production run of MBMT MD. Two reweighted probability distributions at pressures, $P^*$ = 0.055 and $P^*$=0.07, clearly show isolated liquid (red) and gas (blue) phases. Co-existing probability distribution is obtained at $T^*$=1.15, $P^*$=0.061. **c**. $<E^*>_{NPT}$, $<V^*>_{NPT}$, $<H^*>_{NPT}$, and $<C_p^*>_{NPT}$ are reweighted as a function of temperature from five production runs of MBMT MD with five different pressures $P^*$ = 0.03, 0.05, 0.08, 0.1, and 0.12. As the pressures become close to the critical point at $T^*$=1.320, $P^*$=0.1288, the heat capacity gets lower, and it becomes difficult to distinguish gas and liquid phases as expected.

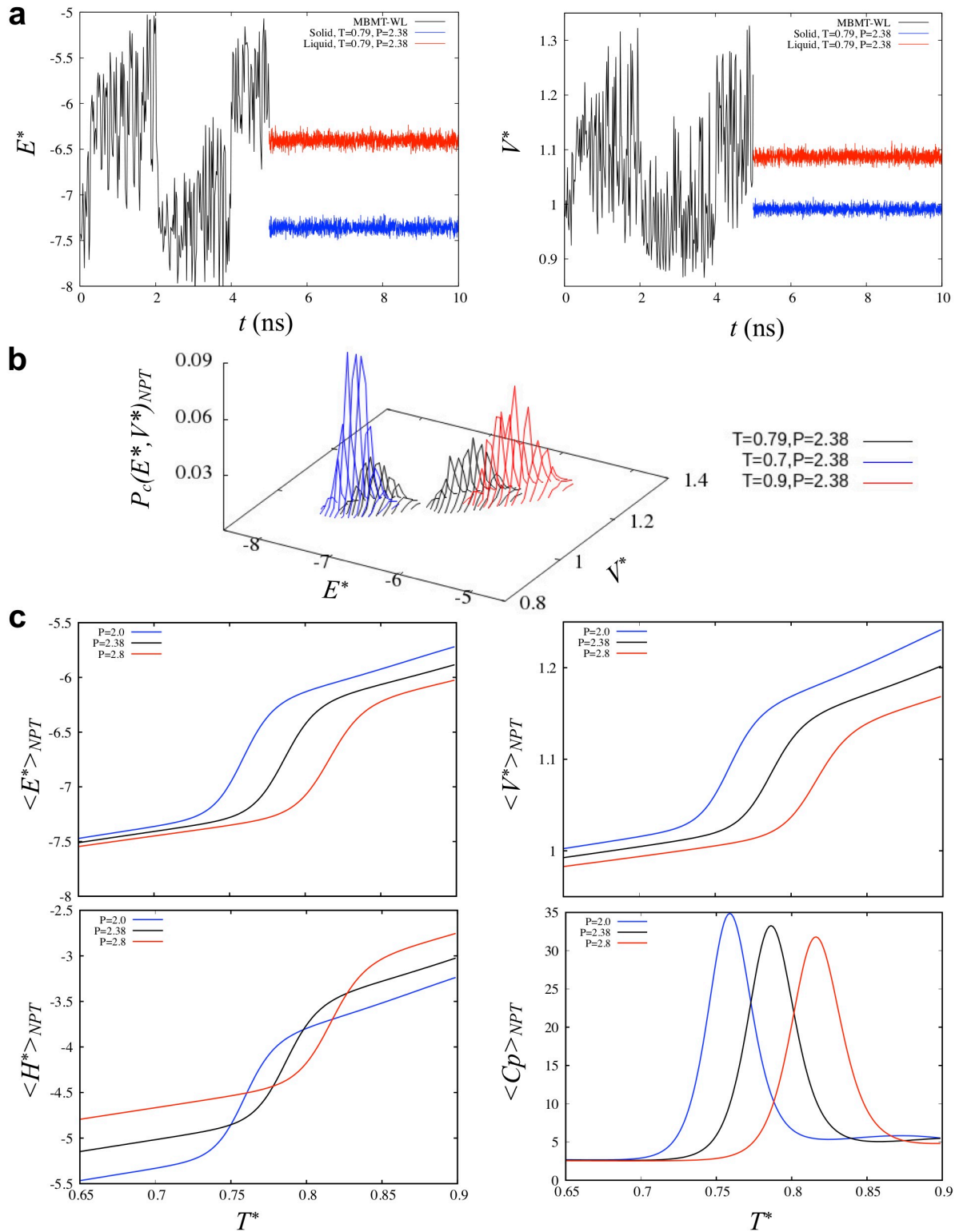

**Figure 4**. **a**. Time evolution of energy per atom ($E^*$) and volume per atom ($V^*$) of solid-liquid system with MBMT ensemble (black: $T^*$=0.72, $P^*$=2.38) and two canonical ensembles ($T^*$=0.79, $P^*$=2.38). One starts from the liquid phase (red), and the other starts from FCC solid phase (blue). MBMT with the WL process can sample both phases. **b**. Probability distributions reweighted from the final production run of MBMT MD. **c**. $<E^*>_{NPT}$, $<V^*>_{NPT}$, $<H^*>_{NPT}$, and $<C_p^*>_{NPT}$ are reweighted as a function of temperature from five production runs of MBMT MD with three different pressures $P^*$ = 2.0, 2.38, and 2.8.

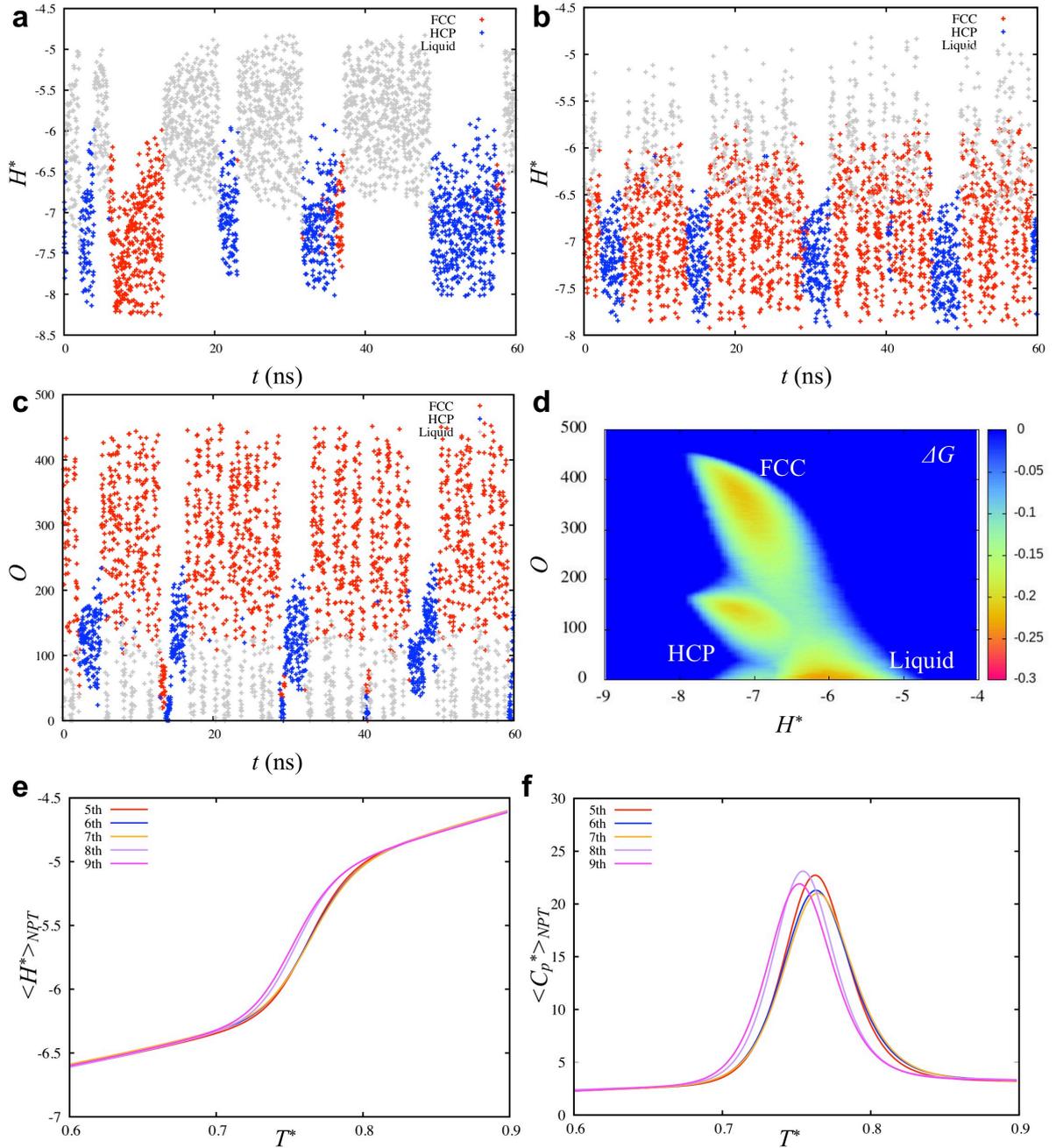

**Figure 5**. **a**. Time evolution of enthalpy per atom ($H^*$) of solid-liquid system with MBMT ensemble at ($T^*$=0.72, $P^*$=2.38). **b**. Time evolution of enthalpy per atom ($H^*$) of solid-liquid system with MOMT ensemble at ($T^*$=0.71, $P^*$=0.02). **c**. The values of bond order parameters utilized in MOMT sampling. **a-c**. The phases (liquid, FCC, and HCP) are labeled by Steinhardt's bond order. **d**. Example of the relative free energy, $8^{th}$ - $\delta H$ ($H^*$,$O$), obtained from MOMT ensemble MD at ($T^*$=0.71, $P^*$=0.02), showing the location of each phase labeled from panel **c**. **e-f**. Examples of $<H^*>_{NPT}$, and $<C_p^*>_{NPT}$ reweighted from five production runs of MOMT MD at $P^*$=1.0.

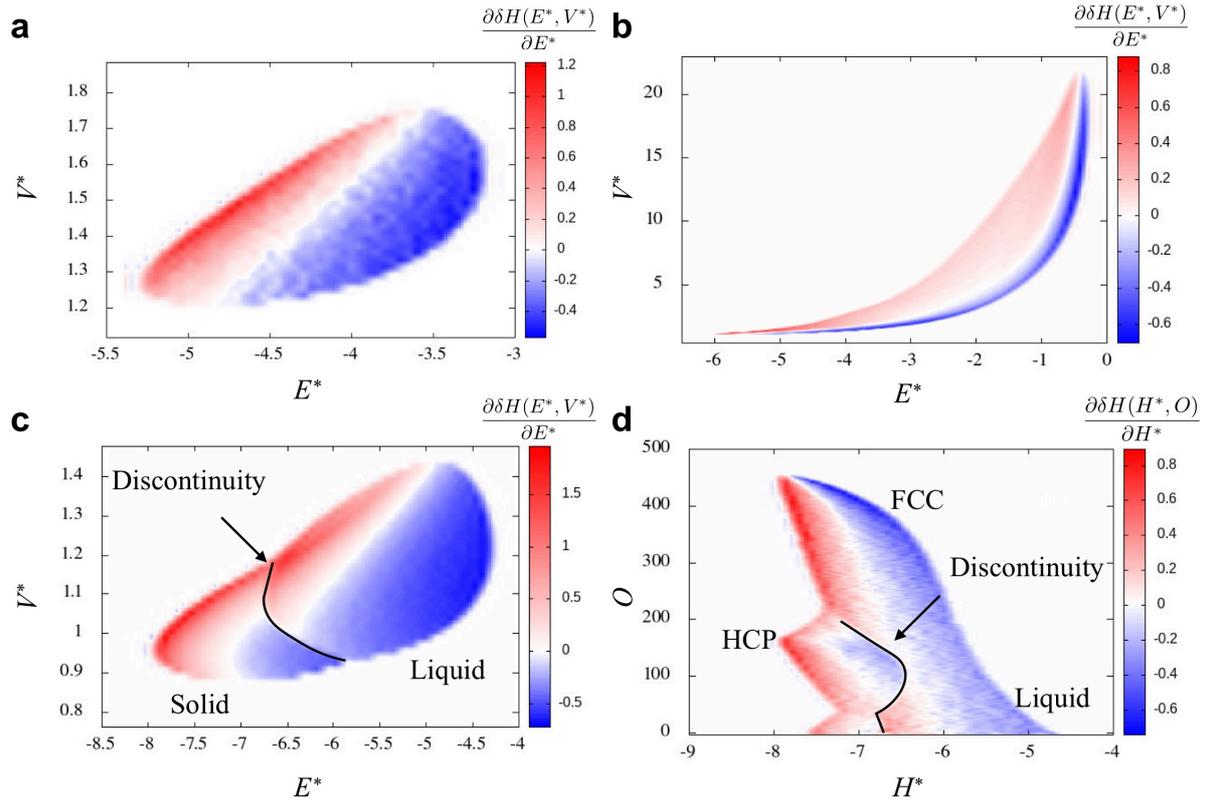

**Figure 6**. **a**. Time evolution of energy per atom ($H^*$) of solid-liquid system with MBMT ensemble at ($T^*$=0.72, $P^*$=2.38). **b**. Time evolution of energy per atom ($H^*$) of solid-liquid system with MOMT ensemble at ($T^*$=0.71, $P^*$=0.02). **c**. The values of bond order parameters utilized in MOMT sampling. **a-c**. The phases (liquid, FCC, and HCP) are labeled by Steinhardt's bond order. **d**. Example of the relative free energy, $8^{th}$ - $\delta H$ ($H^*$, $O$), obtained from MOMT ensemble MD at ($T^*$=0.71, $P^*$=0.02), showing the location of each phase labeled from panel **c**. **e-f**. Examples of $<H^*>_{NPT}$ and $<C_p^*>_{NPT}$ reweighted from five production runs of MOMT MD at $P^*$=1.0.

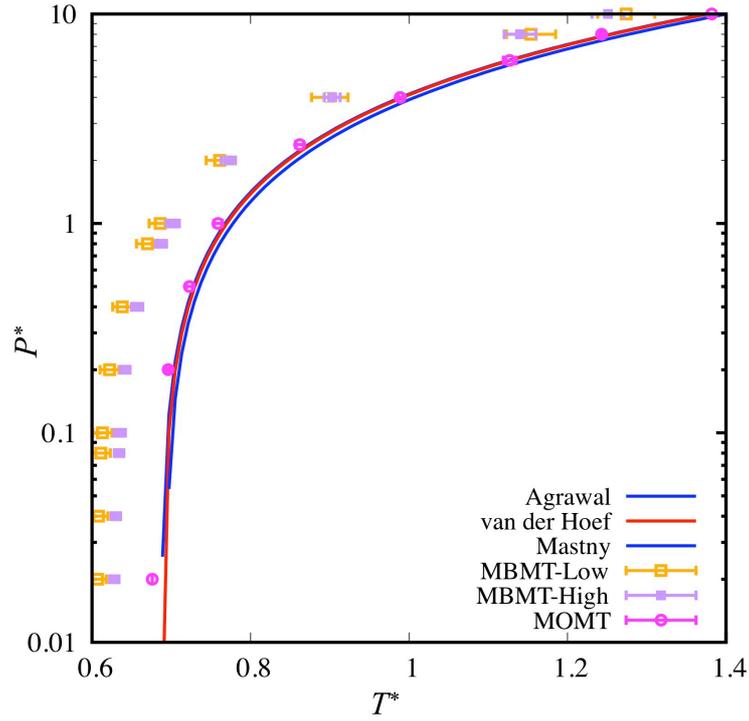

**Figure 7**. Transition temperatures were obtained from the reweighted heat capacity from five MBMT and MOMT MD production runs with error bars. Lines indicate the EOS from eq (27). Agrawal[23]: $A$=-7.19 and $B$=-3.028; van der Hoef[24]: $A$=-7.2866 and $B$=-2.9895; Mastny[25]: $A$ = -8.2269 and $B$ = -2.398. MBMT with two different conditions (Low: $T_0^*$ =0.71, $P_0^*$ = 2.38; High: $T_0^*$ =0.95, $P_0^*$ = 3.00). The obtained transition temperatures from MOMT MD are well-matched with the reference EOS, while MBMT underestimates the transition temperatures.